\documentclass[aps,twocolumn,pra,superscriptaddress,nofootinbib,amsmath,amssymb,floats,floatfix,english]{revtex4-1}
\usepackage{polski}
\usepackage[utf8]{inputenc}
\usepackage[T1]{fontenc}
\usepackage{amsmath}
\usepackage{amssymb}
\usepackage{ifthen}
\usepackage{graphicx}
\usepackage{times}
\usepackage{babel}
\usepackage{color}
\usepackage{bm}
\usepackage{pstricks}

\usepackage{ulem}

\usepackage{lmodern}
\usepackage{ogonek}

\DeclareMathSizes{7}{7}{5}{4} 
\DeclareMathSizes{6}{6}{3}{2} 
\DeclareMathSizes{5}{5}{3}{2} 

\DeclareMathSizes{5}{5}{2}{2}

\DeclareMathSizes{4}{4}{2}{2} 

\newcommand{\Mbig}[1]{\text{\Large${#1}$}}

\newcommand{\ifis}[2]{
\ifthenelse{\equal{#1}{}}{}{#2}
}

\newcommand{\op}[1]{\widehat{#1}}
\newcommand{\dagop}[1]{\widehat{#1}^{\dagger}}
\newcommand{\bo}[1]{{\mathbf{#1}}}
\newcommand{\mc}[1]{{\mathcal{#1}}}
\newcommand{\wt}[1]{{\widetilde{#1}}}
\newcommand{\wb}[1]{{\overline{#1}}}

\newcommand{\nonu}{\nonumber}

\newlength{\templength}

\newcommand{\ifsplit}[1]{\left\{\begin{array}{c@{\qquad}l}#1\end{array}\right.}

\newenvironment{eq}{\begin{equation}}{\end{equation}}

\newcommand{\eqn}[1]{(\ref{#1})}

\renewcommand{\eq}[2]{\begin{equation}\label{#1}#2\end{equation}}
\newcommand{\eqs}[2]{\begin{subequations}\label{#1}\begin{eqnarray}#2\end{eqnarray}\end{subequations}}
\newcommand{\eqa}[2]{\begin{eqnarray}\label{#1}#2\end{eqnarray}}

\usepackage{lmodern}

\newcommand{\ve}{\varepsilon}

\begin{document}
\title{Continuum of classical-field ensembles from canonical to grand canonical and the onset of their equivalence}

\author{J. Pietraszewicz}
\affiliation{Institute of Physics, Polish Academy of Sciences, Aleja Lotnik\'ow
32/46, 02-668 Warsaw, Poland}
\email{pietras@ifpan.edu.pl}

\author{E. Witkowska}
\affiliation{Institute of Physics, Polish Academy of Sciences, Aleja Lotnik\'ow
32/46, 02-668 Warsaw, Poland}

\author{P. Deuar}
\affiliation{Institute of Physics, Polish Academy of Sciences, Aleja Lotnik\'ow
32/46, 02-668 Warsaw, Poland}

\date{\today}

\begin{abstract}
The canonical and grand-canonical ensembles are  two usual marginal cases for ultracold Bose gases,  
but real collections of experimental runs commonly have intermediate properties. 
Here we study the continuum of intermediate cases, and look into the appearance of ensemble equivalence as interaction rises for mesoscopic 1d systems.  
We demonstrate how at sufficient interaction strength the distributions of condensate and excited atoms become practically identical regardless of the ensemble used. 
Importantly, we find that features that are fragile  in the ideal gas and appear only in a strict canonical ensemble can become robust in all ensembles when interactions become strong. 
As evidence, the steep cliff in the distribution of the number of excited atoms is preserved.
To make this study, a straightforward approach for generating canonical and intermediate classical field ensembles using a modified stochastic Gross-Pitaevskii equation (SGPE) is developed.
\end{abstract}
\pacs{}

\maketitle 
\mbox{}

\section{Introduction }
\label{INTRO}

While the canonical and grand canonical ensembles are two dominant ways to describe thermal ultracold Bose gases, ensembles with intermediate fluctuations of particle number are more typical in practice. 
The difference can matter a lot for ultracold experiments because they take place in a mesoscopic regime where such fluctuations are well resolved.

Intermediate ensembles are also theoretically interesting in their own right.
In typical thermodynamic systems without excessively long-range interactions or correlations, the different statistical ensembles are known to give the same result for intensive thermodynamic quantities in the limit of a large system  --- ensemble equivalence \cite{Kocharovsky16,Touchette15,Touchette11,Squartini15}.  However, in a flurry of activity some years ago \cite{Navez97,Weiss97,Grossmann96,Wilkens97,Politzer96,Gajda97,Grossmann97,Giorgini98,Kocharovsky00a,Kocharovsky00b,Xiong01,Xiong02,Yukalov05a,Yukalov05b} it was found that the ideal Bose gas at ultracold temperatures, does not behave this way. Not only do its fluctuations become extremely large in the vicinity of the critical temperature, but the result depends on the ensemble that is used (canonical/grand canonical/microcanonical), even in the thermodynamic limit $N\to\infty$. Ref.~\cite{Kocharovsky06} gives an extensive review. 

Now, with interactions, often even weak ones, it is thought that 
equivalence between ensembles is restored because the interactions energetically suppress any excessive number fluctuations \cite{Wilkens97}. This matter has been, and continues to be, widely debated \cite{Wilkens97,Giorgini98,Idziaszek99,Kocharovsky00a,Yukalov05a,Yukalov05b,Kocharovsky06,Svidzinsky06,Heller13,Cockburn11a,Touchette11,Squartini15,Tarasov15}. Refs.~\cite{Kocharovsky16} and \cite{Touchette15}  explain the current understanding. 
Ensemble equivalence or the thermodynamic limit is often invoked to justify the use of the most convenient ensemble  in calculations.
Hence, the details of how ensemble equivalence is imposed by interactions and what happens in mesoscopic systems are of much interest for practical applications as well as for a theoretical understanding. 

In experiment, the situation is that
repeated runs usually produce a set of states that correspond to something intermediate between a CE and a GCE. The evolution of a single realization conserves particle number when particle loss is neglected, motivating many CE theoretical treatments. However, quite strong fluctuations in total atom number \emph{between} runs are the norm. 
For example, \cite{JaskulaPhD} reports standard deviations $\delta N/N$ of about 20\% or even 35\% with a less optimized system, and a recent study \cite{Kristensen16} about 10\%.
An ensemble with controlled  atom number fluctuations is what best describes an actual set of experimental runs. 
Preferably, one would like an external parameter to match the degree of fluctuation to empirical observations.
A further recent development in this regard are photonic BECs, because the size of the particle reservoir with which they are in contact can be varied to experimentally study the crossover between the GCE and CE in a controlled way \cite{Weiss16}. Experiments with photonic BECs have also been able to measure the distribution $P(N_0)$ directly \cite{Schmitt14}.

In this paper we demonstrate what happens between the CE and GCE in the uniform one-dimensional gas and develop a convenient method to generate intermediate ensembles.
The most adaptable technique to describe degenerate thermal interacting gases are ensembles of classical wave fields (``c-fields''). They are often the only way to gain quantitative access to many quantities in the regime with non-perturbative fluctuations \cite{Brewczyk07,Blakie08,FINESS-Book}. The standard methods developed to date generate only a grand canonical ensemble (GCE) \cite{Stoof99,Duine01,Gardiner02,Gardiner03,Proukakis08,FINESS-Book-Cockburn,Bradley14,Pietraszewicz15}, canonical ensemble (CE) \cite{Witkowska10,Sinatra08,Heller09,Rooney12} or microcanonical ensemble (MCE) \cite{Goral01,Davis01a,Berloff02,Davis03,FINESS-Book-Davis,Brewczyk07,Blakie08,Karpiuk10,FINESS-Book-Brewczyk} of classical fields!
We develop an approach based on the stochastic projected Gross-Pitaevskii equation (SPGPE) \cite{Stoof99,Duine01,Gardiner02,Gardiner03,Proukakis08,FINESS-Book-Cockburn,Bradley14} that readily generates ensembles across the entire continuum from CE to GCE. These transitional ensembles are parametrized by $\sigma$, which determines the standard deviation of the total atom number $N$. Additionally, our method gives more convenient access to the canonical ensemble 
than methods that hardwire exact number conservation into the system such as \cite{Goral01,Davis01a,Berloff02,Davis03,FINESS-Book-Davis,Brewczyk07,Blakie08,Karpiuk10,FINESS-Book-Brewczyk,Witkowska10}.

We will also pay attention to an interesting phenomenon in the CE that has not been extensively investigated.
Namely, the appearance of a ``cliff'' in the distribution of the number of excited particles.
This occurs at relatively high temperatures $T\sim T_c$, when the constraint on $N$ is lower than the number of excited particles suggested by the Bose-Einstein distribution for each mode. Evidence of this feature has been seen in both the ideal \cite{Grossmann96,Wilkens97,Weiss97,Svidzinsky06,Witkowska09,Witkowska10,Tarasov15} and 1d interacting gas in the CE \cite{Carusotto03c,Bezett09b,Bisset09b,Cockburn11a,Bienias11a}. 
No investigation has been made in the GCE  with interactions. It is interesting to find out how robust this phenomenon is to a breaking of the extreme constraint on $N$ that occurs in the CE. 

Prior to that, we derive and describe the SPGPE method for transitional and canonical ensembles in Sec.~\ref{DER}. We benchmark it on the ideal gas in Sec.~\ref{IDGAS}, calculate the distributions and fluctuations in the interacting gas in Sec.~\ref{INT} and look into ensemble equivalence and the ``cliff'' in Sec.~\ref{EQUIV}.

\section{Stochastic method for canonical and transitional ensembles}
\label{DER}

\subsection{The system and its c-field description}
We will consider a single-species gas of contact-interacting bosons. With the Bose field $\op{\Psi}(\bo{x})$, the Hamiltonian is written
\eq{H}{
\op{H} = \int d^d\bo{x}\ \dagop{\Psi}(\bo{x})\left[H^{\rm sp}+\frac{g}{2}\,\dagop{\Psi}(\bo{x})\op{\Psi}(\bo{x})\right]\op{\Psi}(\bo{x})
}
in $d$ dimensions. 
The contact interaction strength is $g$, and $H^{\rm sp}$ is the single particle energy:
\eq{Hsp}{
H^{\rm sp} = -\frac{\hbar^2}{2m}\nabla^2 + V(\bo{x}).
}
On a discretized spatial lattice with small volume per lattice point $dV$, as often used for calculations, \eqn{H} becomes
\eq{Hdx}{
\op{H} \to dV \sum_{\bo{x}} \dagop{\Psi}_{\bo{x}}\left[\sum_{\bo{y}}H^{\rm sp}_{\bo{x}\bo{y}}\op{\Psi}_{\bo{y}}+\frac{g}{2}\dagop{\Psi}_{\bo{x}}\op{\Psi}_{\bo{x}}\op{\Psi}_{\bo{x}}\right].
}
The Hermitian nature of $H^{\rm sp}$ implies 
\eq{Hsp-Herm}{
\left(H^{\rm sp}_{\bo{x}\bo{y}}\right)^* = H^{\rm sp}_{\bo{y}\bo{x}}.
}
We will use this discretized representation interchangeably with the continuous one, according to convenience.

In a minimalist view: the c-field (classical wave field) description boils down largely to an assumption that the relevant behavior of the system is captured by the highly 
occupied modes, while those with $\mc{O}(1)$ occupation or less can be neglected. 
Two complementary reviews of the c-field approach are \cite{Brewczyk07} and \cite{Blakie08}.
We can write the quantum Bose field in terms of orthogonal modes labeled $j$ with mode functions $\phi_j(\bo{x})$ normalized to unity and annihilation operators $\op{a}_j$:
\eq{psiQ}{
\op{\Psi}(\bo{x}) = \sum_j \phi_j(\bo{x})\,\op{a}_j.
}
Then, the c-field approximation corresponds to 
\eq{cfield}{
\op{\Psi}(\bo{x}) \to \left\{\ \psi(\bo{x}) =  \sum_{j\in\mc{C}} \phi_j(\bo{x})\alpha_j \right\}\ ,
}
where $\mc{C}$ is the subspace of high-occupied modes and $\alpha_j$ are complex values that approximate the $\op{a}_j$. The $\{\ \dots\ \}$ indicates that the quantum operator $\op{\Psi}$ is in general going to be described by an ensemble. The numbers $\alpha_j$ will differ among different elements of the ensemble. The subspace $\mc{C}$ is generally chosen \emph{a priori} and specified by an energy cutoff $E_{\rm cut}$, such that all single particle modes with energies below this cutoff are included in $\mc{C}$ and all above excluded. This is 
the most consistent choice for systems that lie close to thermal equilibrium, since occupations will decrease monotonically with energy. A recent detailed study of a broadly applicable cutoff choice for ultracold interacting gases is \cite{Pietraszewicz17}.

The c-field Hamiltonian for the low-energy part of the system  takes the form:
\eq{Hc}{
E(\psi) = \int d^d\bo{x}\ \psi(\bo{x})^*\left[H^{\rm sp}+\frac{g}{2}|\psi(\bo{x})|^2\,\right]\psi(\bo{x})
}
and the number of particles is
\eq{N}{
N(\psi) = \int d^d\bo{x}\ |\psi(\bo{x})|^2.
}

The distribution of $\psi(\bo{x})$ is  then written as
\eq{PGCE}{
P(\psi) \propto \left\{\begin{array}{c} 
 \exp\left(-\,\frac{E(\psi) - \mu N(\psi)}{k_BT}\right) \quad \text{in the GCE} \\
 \exp\left(-\,\frac{E(\psi)}{k_BT} \right) \quad \text{in the CE.}  \end{array}\right.
}

We will use $\hbar=m=k_B=1$ units in what follows.

\subsection{Generation of ensembles}

The two most widespread approaches to produce a c-field ensemble for interacting particles involve generating samples from the ergodic time-evolution of an initial state using the Gross-Pitaevskii equation (GPE) and its variants.
They are: 

(1) Evolution of an initial state using the deterministic but ergodic GPE \cite{Goral01,Berloff02,Brewczyk07,Karpiuk10,FINESS-Book-Brewczyk} or its projected version (PGPE) \cite{Davis01a,Davis03,Blakie08,FINESS-Book-Davis}. This corresponds to isolated Hamiltonian evolution of the classical wave field and produces a microcanonical ensemble (MCE) with number and energy set by the initial state.

(2) Evolving a stochastic Gross-Pitaevskii equation (SGPE) \cite{Stoof99,Duine01,Gardiner02,Proukakis08,FINESS-Book-Cockburn} or its more general projected version (SPGPE) \cite{Gardiner03,Bradley14}, which corresponds to a model where the above-cutoff modes that are excluded from $\mc{C}$ are approximated as a particle and energy bath. This produces a grand canonical ensemble (GCE) with chemical potential $\mu$ and temperature $T$ set externally.

Regarding the canonical ensemble, several methods to generate an interacting  classical wave field ensemble have been proposed:
\begin{enumerate}
\item A Metropolis algorithm for generating samples with a CE probability. 
It involves a discrete random walk taken with steps that conserve particle number \cite{Witkowska10}. It has been used in several studies since \cite{Witkowska11,Karpiuk12,Bienias11a,Pawlowski13,Lang14,Karpiuk15,Gawryluk15,Pawlowski15,Pietraszewicz15,Gawryluk17}, and is also easily adapted to the GCE \cite{Pietraszewicz15,Pietraszewicz17}.
\item A particle number filter applied to grand canonical ensembles to obtain a CE, though this is a wasteful procedure.
\item The noise modifications and projections laid out in \cite{Heller13} constitute another method.
\item Rooney \textsl{et al.} found that an SPGPE with no particle exchange terms produces a canonical ensemble if the exotic scattering terms are included  \cite{Rooney12}.
\item Another approximate approach that can be very accurate under the right conditions used a Bogoliubov description for excited atoms supplanted with as many condensate atoms as required to match the total assumed atom number \cite{Sinatra07}. 
\end{enumerate}

The above canonical ensemble approaches are 
not always easily done, especially when one wants to have e.g. a set magnetization in spinor or multi-component condensates. For example, we have found that ensuring the right conservation law while preserving detailed balance becomes very tricky with Metropolis for multicomponent gases.
It is known also that the implementation of the scattering-only SGPE is a nontrivial endeavor even for single component gases \cite{Blakie08,Rooney12}. 

In this paper 
we derive an alternative approach that extends the SPGPE to incorporate a controllable number filter. It restricts the
evolution to the vicinity of the CE and avoids the waste of discarding realizations. Moreover, it gives access to natural
intermediate ensembles between the marginal CE and GCE.

\subsection{The SPGPE}

The SPGPE is a flexible way to generate the grand canonical ensembles of classical wave fields given by  \eqn{PGCE} \cite{Blakie08,Staliunas06}.
It has been described in detail in \cite{Duine01,Gardiner03,Proukakis08} and benchmarked in \cite{Cockburn11a,Cockburn11c}. 

In general, one works in a projected subspace $\mc{C}$.  It is imposed  by acting with a projector $\mc{P}$ onto spatially dependent fields $f(\bo{x})$ such that
$\mc{P}f(\bo{x})$ lies wholly within $\mc{C}$. 
This allows one to work on a simple spatial grid $\bo{x}$ while restricting the basis in any desired way. 
On the spatial lattice,  this becomes
\eq{Pf}{
\mc{P}f(\bo{x}) = \sum_{\bo{y}}\mc{P}_{\bo{x}\bo{y}} f_{\bo{y}}.
}
An explicit form of the matrix elements is
\eq{P}{
\mc{P}_{\bo{x}\bo{y}} = dV \sum_{j\in\mc{C}} \phi_j(\bo{x}) \phi^*_j(\bo{y}).
}
The projector $\mc{P}$ fulfills the usual properties:
\eq{Pprop}{
\mc{P}\mc{P} = \mc{P},\qquad
\mc{P}^{\dagger} = \left(\mc{P}^T\right)^* = \mc{P}, \qquad
\mc{P}^T = \mc{P}^*.
}
The simplified case of a plane wave basis with cutoff set implicitly by the lattice corresponds to setting $\mc{P}\to1$.

The time evolution of $\psi(\bo{x})$ is governed by the SPGPE
\eqa{SPGPE}{
\frac{\partial\psi(\bo{x})}{\partial t} &=& \mc{P}\Big\{-(i+\gamma(\bo{x}))\left[H^{\rm sp}-\mu+g|\psi(\bo{x})|^2\right]\psi(\bo{x})  \nonu\\
&& + \sqrt{2T
\gamma(\bo{x})}\ \eta(\bo{x},t)\Big\}.
}
It corresponds to coupling the c-field to a thermal bath at temperature $T$ and chemical potential $\mu$.
The dimensionless positive coupling strength is $\gamma(\bo{x})$, which 
is commonly taken to be constant in space. Such an assumption is
certainly a convenience if one is primarily interested in the long-time ensemble, rather than the transient dynamics. Physically justified values are usually small ($\gamma\ll1$).
$\eta$($\bo{x},t$) is a complex white noise field independent at each spatial position and time, with zero mean, and variance:
\eq{etaeta}{
\langle\eta(\bo{x},t)^*\eta(\bo{x}',t')\rangle = \delta^d(\bo{x}-\bo{x}')\delta(t-t').
}
The equation \eqn{SPGPE} changes only that part of the field $\psi$ that has support in the c-field subspace $\mc{C}$. 
To be self-consistent and physically sensible we need the initial state $\psi_0(\bo{x})$ to be fully in this subspace, i.e. 
\eq{Qpsi=0}{
(\mc{P}-1)\psi_0(\bo{x})=0.
}

To obtain a GCE  one evolves the equation until transients related to the initial state have died off, and only thermally activated fluctuations remain. Let us call this time $t_*$. The choice of initial state is, in principle, irrelevant, although it may affect the length of time needed to reach the thermally activated regime. Starting from vacuum $\psi(\bo{x},0)=0$ is a common choice. Independent samples of the distribution can then be obtained from values of the field $\psi(\bo{x},t)$ sufficiently well spaced in time after $t_*$. This is reminiscent in many ways of the procedure with the Metropolis method, except that 
all updates are accepted, and given explicitly by the noise term.
Alternatively, one can simply evolve from the same initial state to $t=t_*$ but using a different noise realization each time, and the fields $\psi(\bo{x},t_*)$ will be the independent samples of the GCE. The latter approach removes the need to investigate time correlations. It also simplifies the determination of $t_*$ because ensemble-averaged quantities can be tracked for a number of times leading up to $t\gtrsim t_*$, to verify 
when the stationary ergodic ensemble has been reached. 
 
The system evolves to the GCE distribution \eqn{PGCE} regardless of the details of $\gamma(\bo{x})>0$, which only affects the time $t_*$.

\subsection{One mode and ensemble equivalence }
\label{1MODE}

It is instructive in the beginning to look at the behavior of ensembles in a single mode, $j=0$ say.
We will see that this example encapsulates both the basic physics of how ensemble equivalence is restored by interactions, and suggests a naturally occurring form for the manifold of intermediate ensembles. 

The Hamiltonian \eqn{Hc} in the c-field description can be written as an energy
\eq{H1}{
E(\alpha) = \omega|\alpha|^2 + gc|\alpha|^4,
} 
that depends on the amplitude $\alpha$. The coefficients depend on the shape of the mode function according to
$\omega=\int d^d\bo{x}\,\phi_0(\bo{x})^*H^{\rm sp}\phi_0(\bo{x})$
and
$c=\tfrac{1}{2}\int d^d\bo{x}\,|\phi_0(\bo{x})|^4$, while the  number of atoms is $N(\alpha)=|\alpha|^2$.
According to \eqn{PGCE} the distribution of the states in the GCE is
$
P_{GCE}(\alpha) \propto\ \exp\left\{-\,\frac{(\omega-\mu)|\alpha|^2 + gc|\alpha|^4}{T}\right\}
$ 
with all values of $\alpha$ represented.

For an ideal gas, the above exponent produces a very broad distribution of particle number
\eq{PGCEid}{
P_{GCE}(N) \ \propto\  e^{\frac{(\mu-\omega)N}{T}}.
}
This is the most trivial case of the GCE fluctuation catastrophe and inequivalence of ensembles,
since the fluctuations of $N$ scale as $N$. In fact $\delta N = \sqrt{\langle N^2\rangle - \langle N\rangle^2} = \frac{\omega-\mu}{T} = N$,
so $P_{GCE}(N)$ never approaches the CE behavior of $\delta N=0$, even as $N\to\infty$.

Interactions, however, make the distribution Gaussian:
\begin{subequations}\label{PGCEg-eq}
\eq{PGCEg}{
P_{GCE}(N) \propto\ \exp\left[-\,\frac{\left(N-N_{\rm mid}\right)^2}{2\sigma_{\rm1mode}^2}\right]
}
with
\eq{PGCEgm}{
N_{\rm mid} = \frac{\mu-\omega}{2gc}
}
and
\eq{PGCEgs}{
\sigma_{1\rm mode} = \sqrt{\frac{T}{2gc}}.
}
\end{subequations}
Now we can see that, with the help of the interaction $g$, one can drive the standard deviation of the Gaussian, $\sigma_{1\rm mode}$, to 
smaller values. Eventually, the fluctuations in particle number, $\delta N$,  become $\sigma_{\rm1mode}\propto1/\sqrt{g}$. 
Thus, for large $N$ and $\sigma_{\rm1mode}\ll N_{\rm mid}$, ensemble equivalence is restored in the thermodynamic limit $N\to\infty$  because $\delta N/N \to 0$ like in the CE. 
In terms of $g$, this happens for 
\eq{gequiv}{
g \gtrsim \frac{(\mu-\omega)^2}{2cT}.
}
This example shows the essence of how ensemble equivalence is restored by interactions.

\subsection{Restricting the atom number in the SPGPE via an additional term}
\label{DERIV}
It would be convenient to have an equation that explicitly conserves $N(\psi)$ to a set value $\wb{N}$ but keeps a similar form as the SPGPE \eqn{SPGPE}. 
And indeed -- the one mode toy problem of Sec.~\ref{1MODE} suggests a way: Terms of a similar form to the interaction term should be capable of imposing a Gaussian distribution of $N$ with a width of our choice, while leaving the rest of the system evolution largely unchanged. 
In the limit of a narrow Gaussian distribution around the desired value, we would have effectively a CE distribution.

Consider an additional Gaussian factor to \eqn{PGCE} thus:
\eq{PGCEs}{
P_{\sigma}(\psi) \propto \exp\left\{-\,\frac{\left[E(\psi) - \mu N(\psi)\right]}{T}-\frac{\left(N(\psi)-\wb{N}\right)^2}{2\sigma^2}\right\}.
}
When $\sigma$ becomes smaller than other widths, only fields $\psi(\bo{x})$ with  a number of particles $\wb{N}\pm\sigma$ occur with non-negligible probability. 
In the limit of small $\sigma$ this becomes effectively a CE with $\wb{N}$ particles. 
What terms should be added to the SPGPE to attain this modification?

First note that the exponent of the probability distributions of c-field states contain all the terms of the Hamiltonian, converted to a classical field and scaled. 
Secondly, the deterministic parts of the SPGPE correspond to the classical field simplification of the Heisenberg equations of motion for the quantum field $\op{\Psi}$. i.e. of
$d\op{\Psi}(\bo{x})/dt = -i\left[\op{\Psi}(\bo{x}),\op{H}\right]$. Hence, each term $\op{H}_j$ in the Hamiltonian leads to the c-field version of $-\mc{P}\{(i+\gamma)\left[\op{\Psi}(\bo{x}),\op{H}\right]\}$ in the SPGPE. 
Taken together, these two points suggest that the new term in \eqn{PGCEs} proportional to the c-field version of $(\op{N}-\wb{N})^2$ will generate a term in the stochastic equation  proportional to the c-field version of $[\op{\Psi}(\bo{x}),(\op{N}-\wb{N})^2]$. 
That is, one may expect terms proportional to $\psi(\bo{x})(N(\psi)-\wb{N})$. Let us postulate, then, a modified SPGPE:
\eqa{SPGPEmod}{
\frac{\partial\psi(\bo{x})}{\partial t} &=& \mc{P}\Big\{-(i+\gamma(\bo{x}))\left[H^{\rm sp}-\mu+g|\psi(\bo{x})|^2\right]\psi(\bo{x})  \nonu\\
&& \hspace*{-4em}+ \sqrt{2T
\gamma(\bo{x})}\ \eta(\bo{x},t)    
+ K(\bo{x})\left[N(\psi)-\wb{N}\right]\psi(\bo{x})\ \Big\}
}
with a constant $K$ (possibly space-dependent) to be determined. We will see if and for what value of $K$ the stationary distribution is equal to the desired \eqn{PGCEs}.
Note that we have placed the term inside the projection $\mc{P}$ because the equation should always preserve the property that $\psi(\bo{x})$ has support only in the $\mc{C}$ subspace to have  a consistent c-fields description.
 
The correspondence between stochastic equations, Fokker-Planck equations for the distribution, and stationary states is well known. A detailed explanation can be found e.g. in \cite{StochMech}.
When one has a set of real variables $\vec{v}=\{v\}$ governed by Langevin stochastic equations of the form 
\eq{Langevin}{
\frac{dv}{dt} = \sum_v A_v(\vec{v}) + \sum_{uv} B_{uv}(\vec{v})\,\xi_v(t),
}
with real noises of zero mean and correlations
\eq{xi}{
\langle\xi_u(t)\xi_v(t')\rangle = \delta_{uv}\delta(t-t'),
}
then it is a realization of the following Fokker-Planck equation (FPE) for the probability distribution $P(\vec{v})$ of the variables:
\eq{FPE}{
\frac{\partial P(\vec{v})}{\partial t} = \left[-\sum_v\frac{\partial}{\partial v} A_v(\vec{v}) + \frac{1}{2}\sum_{uv} \frac{\partial^2}{\partial u\,\partial v} D_{uv}(\vec{v}) \right]P(\vec{v}).
}
Summing is over all variables in $\vec{v}$, and derivatives act on all factors to the right.
The diffusion matrix $D$ is given by elements
\eq{FPE-D}{
D_{uv} = \sum_{v'} B_{uv'} B_{vv'}.
}

The desired stationary distribution is \eqn{PGCEs}, so that we want to impose 
\eq{impose}{
\frac{\partial P(\vec{v})}{\partial t} = 0
}
when the substitution 
\eq{sub}{
P(\vec{v}) = P_{\sigma}(\psi)
}
is made in the FPE \eqn{FPE}.

Consider the system on a numerical lattice as in \eqn{Hdx}, so that the set of variables in $\vec{v}$ consists of
the real and imaginary components of $\psi$ at each point $\bo{x}$, i.e.  $\psi^R_{\bo{x}}={\rm Re}[\psi_{\bo{x}}]$ and $\psi^I_{\bo{x}}={\rm Im}[\psi_{\bo{x}}]$, respectively.
The equation \eqn{SPGPEmod} can be rewritten in the general $A,B,D$ notation of \eqn{Langevin}-\eqn{FPE-D}
using the coefficients
\eqa{A}{
A_{\psi^R_{\bo{x}}}&+&iA_{\psi^I_{\bo{x}}} = \left(N(\psi)-\wb{N}\right) \sum_{\bo{y}}\mc{P}_{\bo{x}\bo{y}}K_{\bo{y}}\psi_{\bo{y}}\\
&&\hspace*{-1em}
- \sum_{\bo{y}}\mc{P}_{\bo{x}\bo{y}}(i+\gamma_{\bo{y}})\left\{
\sum_{\bo{z}}H^{\rm sp}_{\bo{y}\bo{z}}\psi_{\bo{z}} +\left[\,g|\psi_{\bo{y}}|^2-\mu \right]\psi_{\bo{y}}
\right\}\nonu
}
and 	
\eqs{D}{
D_{\psi^R_{\bo{x}},\psi^R_{\bo{y}}} =&  D_{\psi^I_{\bo{x}},\psi^I_{\bo{y}}} &= 
\frac{T}{dV}\ {\rm Re}\left[\sum_{\bo{z}}\mc{P}_{\bo{x}\bo{z}}\gamma_{\bo{z}}\mc{P}_{\bo{z}\bo{y}}^*\right]\\
D_{\psi^R_{\bo{x}},\psi^I_{\bo{y}}} =&  -D_{\psi^I_{\bo{x}},\psi^R_{\bo{y}}} &= 
-\frac{T}{dV}\ {\rm Im}\left[\sum_{\bo{z}}\mc{P}_{\bo{x}\bo{z}}\gamma_{\bo{z}}\mc{P}_{\bo{z}\bo{y}}^*\right]\quad 
}

In the interest of clarity we will assume for now a constant $\gamma$ and $K$:
\eq{gamconst}{
\gamma(\bo{x})\to\gamma, \qquad K(\bo{x})\to K
} 
and report the more general result from Appendix~\ref{APP:GAMMA} at the end of this section.

Substituting  \eqn{sub}-\eqn{gamconst} into \eqn{FPE} leads (after much algebra) to the FPE:
\begin{widetext}
\eqa{FPE-1}{
\frac{\partial P_{\sigma}}{\partial t} &=& 2P_{\sigma}\left[{\rm Re}[K] + \frac{\gamma T}{\sigma^2}\right]
\left\{
dV\sum_{\bo{x}\bo{y}}\mc{P}_{\bo{x}\bo{y}}\psi_{\bo{y}}\psi_{\bo{x}}^*\left[\left(N(\psi)-\wb{N}\right)\left(\frac{N(\psi)-\wb{N}}{\sigma^2}-\frac{\mu}{T}\right)-1\right]-\left(N(\psi)-\wb{N}\right)\sum_{\bo{x}}\mc{P}_{\bo{x}\bo{x}}\right\}\nonu\\
&&+\frac{P_{\sigma}dV}{T}\left(N(\psi)-\wb{N}\right)\sum_{\bo{x}\bo{y}}\mc{P}_{\bo{x}\bo{y}}\psi_{\bo{x}}^*\left\{
\left[g\psi_{\bo{y}}|\psi_{\bo{y}}|^2+\sum_{\bo{z}}H^{\rm sp}_{\bo{y}\bo{z}}\psi_{\bo{z}}\right]
\left[K^*+\frac{\gamma T}{\sigma^2}-i\frac{T}{\sigma^2}\right]
\right\}+{\rm c.c.}
}
\end{widetext}
The first line of \eqn{FPE-1} can easily be made zero with an appropriate choice of $K$, but even then, the second line still remains potentially troublesome. 
However, note that the properties of the equation \eqn{SPGPEmod} and initial state \eqn{Qpsi=0} ensure that the field $\psi$ stays in the c-field subspace in the overall model. 
Then, 
\eq{Pxypsix}{
\sum_{\bo{x}}\mc{P}_{\bo{x}\bo{y}}\psi_{\bo{x}}^*=\psi_{\bo{y}}^*.
}
Using this, \eqn{FPE-1} becomes
\eqa{FPE-2}{
&\dfrac{\partial P_{\sigma}}{\partial t} &= 2P_{\sigma}\left[{\rm Re}[K] + \frac{\gamma T}{\sigma^2}\right]\times\\
&&\hspace*{-1.5em}\Bigg\{\frac{N(\psi)-\wb{N}}{T}\left[dV\sum_{\bo{x}}\left(g|\psi_{\bo{x}}|^4+\sum_{\bo{y}}\psi_{\bo{x}}^*H^{\rm sp}_{\bo{x}\bo{y}}\psi_{\bo{y}}\right)-\mu N(\psi)\right]\nonu\\
&&+\frac{N(\psi)\left(N(\psi)-\wb{N}\right)^2}{\sigma^2} - (N(\psi)-\wb{N})\sum_{\bo{x}}\mc{P}_{\bo{x}\bo{x}} -N(\psi)
\Bigg\}.\nonu
}
The following choice of prefactor on the first line: 
\eq{K=}{
K = -\frac{\gamma T}{\sigma^2}.
}
makes the distribution \eqn{PGCEs} stationary. This is exactly what we required. 
The final equation to simulate is then simply:
\begin{widetext}
\eq{SPGPEs}{
\frac{\partial\psi(\bo{x})}{\partial t} = \mc{P}\left\{
-(i+\gamma)\left[H^{\rm sp}-\mu+g|\psi(\bo{x})|^2\right]\psi(\bo{x})  
-\frac{\gamma T}{\sigma^2}
\left(N(\psi)-\wb{N}\right)\psi(\bo{x})
+\sqrt{2\gamma T}\ \eta(\bo{x},t)    
\right\}.
}
\end{widetext}

Appendix~\ref{APP:GAMMA} explains what happens when we relax condition \eqn{gamconst}
allowing the coupling strength $\gamma(\bo{x})$ to be spatially dependent.

The condition needed to obtain a well behaved equation is that
$\gamma(\bo{x})$ varies slowly in the region of space around $\bo{x}$ compared to $\mc{P}_{\bo{x}\bo{y}}$.
i.e. 
\eq{slow}{
\mc{P}_{\bo{x}\bo{y}}\gamma_{\bo{x}} \approx \mc{P}_{\bo{x}\bo{y}}\gamma_{\bo{y}}.
} 
It is met in most realistic cases.
 Then, the choice \eqn{K=} turns out to generalize to 
\eq{K=gen}{
K(\bo{x}) = -\frac{\gamma(\bo{x}) T}{\sigma^2}.
}
and the modified SPGPE equation that keeps \eqn{PGCEs} stationary is
\begin{widetext}
\eq{SPGPEs-gen}{
\frac{\partial\psi(\bo{x})}{\partial t} = \mc{P}\left\{
-(i+\gamma(\bo{x}))\left[H^{\rm sp}-\mu+g|\psi(\bo{x})|^2\right]\psi(\bo{x})  
-\frac{\gamma(\bo{x}) T}{\sigma^2}
\left(N(\psi)-\wb{N}\right)\psi(\bo{x})
+\sqrt{2\gamma(\bo{x}) T}\ \eta(\bo{x},t)    
\right\}.
}
\end{widetext}
This is just the usual SPGPE with \textbf{one extra term}.

The most likely place for a breakdown of the condition \eqn{slow} in typical problems is in the low density tails of a trapped system that is described using harmonic oscillator modes.  Away from the main cloud, near the energy cutoff, only very low wavelength parts of modes $\phi_j(\bo{x})$ are present, and then  $\mc{P}_{\bo{x}\bo{y}}$ may vary on comparable scales to $\gamma(\bo{x})$.

The special but common case of an unprojected ``plain'' SGPE, where the only projection is an implicit one imposed by the numerical lattice ($\mc{P}_{\bo{x}\bo{y}}=\delta_{\bo{x}\bo{y}}$) also uses \eqn{SPGPEs-gen} with $\mc{P}\to1$, and without the need for the slowly varying condition on $\gamma$ that is \eqn{slow}.

To conclude this derivation,
one can safely say that the canonical ensemble has been achieved for small $\sigma$ when all relevant observable quantities cease to depend on $\sigma$ in any significant way.

The equations \eqn{SPGPEs} and \eqn{SPGPEs-gen} are a convenient way by which one can generate the CE. Both equations are 
stable, straightforward to integrate, and require fewer numerical tweaks than a Metropolis algorithm. 
We only need to set $\gamma$, which can be chosen over a wide range without ill effect, when the purpose is to generate a stationary ensemble.
Furthermore there is no wastage due to particle number filtering.

The equations \eqn{SPGPEs} and \eqn{SPGPEs-gen}, 
can also be used to produce the dynamics of a canonical ensemble,
but then one should determine a correct value and spatial dependence of the reservoir coupling $\gamma(\bo{x})$. 
The question of how realistic the physical model is remains somewhat open, since
the system corresponds to having a low-energy part of the field that exchanges only energy but not particles with the high energy components that are treated as a bath.
Nevertheless, such a model has been discussed in some detail in the context of a scattering-only SPGPE \cite{Rooney12}, and may be useful in various situations.

Overall, the computational cost scales the same way as in other treatments based around the SGPE. That is, a very lenient $M\log M$ scaling with the number of points on the computational lattice $M$, regardless of the dimensionality. This makes it convenient for 2d and 3d systems.
The usual limiting factor is the efficiency of a Fourier transform used to evaluate kinetic energy. 
An issue to keep in mind is that very small values of $\sigma$ will shorten the required timestep by virtue of introducing a large gradient. It may also tend to increase the time $t_*$ required to obtain stationarity. 
Some precursors of this were seen at the lowest values of $\sigma$ in our 1d calculations.

\subsection{Transitional distributions  between CE and GCE}
\label{TRANS}

Equation \eqn{SPGPEs-gen} generates the family of ensembles \eqn{PGCEs} as its long time stationary distribution. These span the whole continuum between CE and GCE for interacting systems, with the location on the continuum given by $\sigma$. 

A convenient way to specify distributions intermediate between CE and GCE is through the standard deviation of the atom number fluctuations $\delta N$. This captures the foremost difference between the CE and GCE, and  can be readily matched to experimental data such as in \cite{JaskulaPhD}. 

There are two contributions to $\delta N$: First, the ``natural'' one ($\delta^{GCE} N$) that arises as a result of the interplay of the interaction strength $g$ and the particle bath described by the chemical potential $\mu$. For the single mode this is \eqn{PGCEgs}. Then there is also the externally steered fluctuation $\sigma$. It will not increase fluctuations beyond the natural level, but can decrease them. Hence, we expect that 
\eq{deltaN}{
\delta N \approx \ifsplit{
\sigma &          \text{if}\ \sigma \lesssim \delta^{GCE} N\\
\delta^{GCE}N &   \text{if}\ \sigma \gtrsim \delta^{GCE} N
}
}
The largest values of $\sigma$ do not affect the GCE much. Then, when $\sigma$ becomes small enough to limit the natural fluctuation width, it begins to meaningfully steer the distribution. Finally, when $\sigma$ becomes small enough that observable quantities cease to change, we have reached the CE.
This changeover will be seen later in Fig.~\ref{fig:lowT-dN/N}.

In the ideal gas, for small enough $\sigma$, the center of the Gaussian-like distribution for $N$ can be quite well estimated by 
\eq{Nmu}{
\wb{N}_{\sigma} \  \approx\  \wb{N} +\frac{\mu\sigma^2}{k_BT}.
}
This comes from inspection of \eqn{PGCEs} while omitting the $H^{\rm sp}$ contribution. 
 $\wb{N}_{\sigma}$ converges to $\wb{N}$ in the CE limit. 
For large $\sigma$, \eqn{Nmu} becomes inaccurate because other factors come into play, such as a nontrivial $H^{\rm sp}$ contribution and the fact that the distribution of $N$ is nonzero only for $N>0$.

In a uniform interacting gas in volume $V$, the properties of the Gaussian can also be estimated. 
The energy functional is
$E(\psi) = \ve_{\rm sp} N(\psi) + \frac{g}{2V}\, g^{(2)}(0) N(\psi)^2$, 
where $\ve_{\rm sp}$ is the mean energy per particle from the single-particle Hamiltonian $H_{\rm sp}$, and $g^{(2)}(0)$ is the density-density correlation function. $g^{(2)}(0)$ lies between 1 and 2 in an equilibrium ensemble. Looking first at the natural GCE in \eqn{PGCE}, the Gaussian distribution for $N(\psi)$ is centered at
\eq{natural-mean}{
\wb{N}_{GCE} = \frac{V(\mu-\ve_{\rm sp})}{g^{(2)}(0)g},
}
with a standard deviation
\eq{natural-sd}{
s_{GCE} = \sqrt{\frac{T\, V}{g^{(2)}(0)g}}.
}
For the transition distributions $P_{\sigma}$ of \eqn{PGCEs}, the center of the Gaussian for $N(\psi)$ shifts to
\eq{trans-center}{
\wb{N}_{\sigma} = \frac{\wb{N}+\wb{N}_{GCE}\,\frac{\sigma^2}{s_{GCE}^2}}{1+\frac{\sigma^2}{s_{GCE}^2}},
}
and the standard deviation becomes
\eq{trans-sd}{
s_{\sigma} = \frac{\sigma}{\sqrt{1+\frac{\sigma^2}{s_{GCE}^2}}}.
}
One can see that indeed in the $\sigma\ll s_{GCE}$ limit, both quantities converge to the externally set values of $\wb{N}$ and $\sigma$, while in the opposite $\sigma\gg s_{GCE}$ limit, the natural GCE behavior reasserts itself.

Unplanned behavior can occur if the difference between $\wb{N}$ and the natural $\wb{N}_{GCE}$ is much greater than $\sigma$. In that case, the external constraint $\wb{N}$ and the internal chemical potential $\mu$ work against each other. The result is a relatively narrow distribution that is not centered near $\wb{N}$ but at a weighted average of $\wb{N}$ and $\wb{N}_{GCE}$ given by \eqn{trans-center}. 
The upshot of this for generation of canonical ensembles in general cases is that one should check the actual resulting mean particle number. If it does not closely match $\wb{N}$, then $\mu$ should be modified to bring $\wb{N}_{GCE}$ close to $\wb{N}$.


\section{Ideal gas}
\label{IDGAS}

Let us first check the method on the ideal gas, where exact results are available. The typical observables studied in the context of comparing ensembles are the distributions $P(N_0)$ and $P(N_{\rm ex})$ of the number of atoms in the ground or excited states, as well as their moments. 
An experimental method for measuring fluctuations in the condensate occupation $N_0$ has been proposed in \cite{Bezett09b}.

\begin{figure}
\begin{center}
\includegraphics[width=\columnwidth]{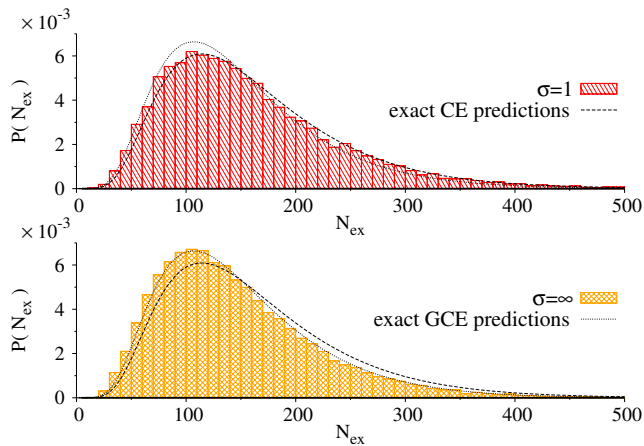}
\end{center}
\caption{
The probability distribution of having $N_{\rm ex}$ excited atoms in a uniform 1d ideal gas 
at a relatively low temperature $T=0.341 T_*$. 
Shown are the GCE (lower, yellow, $\sigma=\infty$) and CE (upper, red, $\sigma=1$) cases.  Dotted and dashed lines correspond to exact classical field 
predictions \eqn{PNex-GCE} and \eqn{PNex-CE}, respectively. They are indiscernible from the full quantum predictions \eqn{PNex-CE-Q} and \eqn{PNex-GCE-Q} for these parameters. 
The target total atom number was fixed at $\overline{N}=500$ in both cases. 
A visible and distinguishable difference between the GCE and CE is confirmed by the histograms.
}
\label{fig:lowT-hist-PNex}
\end{figure}

\subsection{Procedure}
\label{PROC}
We treat here a 1d uniform gas, and the procedure outlined below was applied for both ideal and interacting gases. The chosen basis consists of plane waves $\phi_j(x) = e^{ik_jx}/\sqrt{L}$ defined in a box of length $L$ with periodic boundary conditions. 
Wave vectors are $k_j=2\pi j/L= j\Delta k$ with $j=\{0,\pm1,\pm2,\dots\}$.
We take $L$ to be the computational unit of length in what follows, and only write it out explicitly in a few cases to show scaling. 
The c-field subspace $\mc{C}$ is implemented using a maximum kinetic energy cutoff for the plane waves $E_c=\hbar^2k_c^2/2m$. 

We revisit the regimes that were investigated in the past work of \cite{Witkowska10} (Fig. 1). Namely, we study a similar condensate fraction $n_0=N_0/N$ and distribution of excited atoms. We fixed the target total atom number in the CE at the higher value of $\wb{N}=500$. 

It is convenient to give the temperature scaled with respect to a finite-size characteristic temperature for condensation. In the ideal gas canonical ensemble, 
the occupation of excited modes is given by 
\mbox{$N_k\approx N_k^{\rm BE}=[e^{k^2/2T}-1]^{-1}$},
provided the total number of excited atoms \mbox{$N_{\rm ex} = \sum_{j\neq0} N_{k_j}$} does not reach $\wb{N}$.
Otherwise, it invokes the constraint and mode occupations reduce below $N^{\rm BE}_k$.
To estimate the temperature $T_*$ below which a significant condensate will appear, one can evaluate the simple condition 
$N_{\rm ex}(T_*)=\wb{N}$ using the estimates $N_k^{\rm BE}$.
In our particular case of $\wb{N}=500$, we find 
$T_*=3195/L^2.$
The simplest general estimate comes from considering only the two lowest lying excited states, in which case
$T_*=\pi^2\wb{N}/L^2$. 

The cutoff used for calculations was the recommended value for matching the condensate fraction and $P(N_{\rm ex})$ in a 1d ideal gas in a box in the CE \cite{Witkowska09}
\footnote{This corresponds to $f_c=1.9023$ in the global optimized cutoff notation of \cite{Pietraszewicz15,Pietraszewicz17}, where $k_c=f_c\sqrt{2\pi T}$.}, 
i.e. $k_c^2 = 0.58 T$. 
This leads to a cutoff of $4\Delta k$  for the low temperature $T=0.341\, T_*$ of Sec.~\ref{LOWT} and $k_c=8\Delta k$ for the high temperature $T=1.365\, T_*$ of Sec.~\ref{HIGHT}, like in the work of \cite{Witkowska10}.

The generation of each member of the ensemble proceeds by starting with the vacuum state $\psi_{\bo{x}}(0)=0$ on a numerical lattice with spacing $\Delta x=L/2^7$. Note that the maximum allowable wavevector on this lattice, $k_{\rm max}=\pi/\Delta x$ is much greater than the cutoff $k_c$. This allows us to accurately calculate the interacting evolution, which would otherwise suffer from some small but spurious aliasing and umklapp processes on a lattice with $k_{\rm max}=k_c$. 
The state $\psi_{\bo{x}}(t)$ is then evolved using \eqn{SPGPEs} with a constant value of $\gamma$ 
until a stably randomly fluctuating solution is reached above $t_{*}$.
We used values of $\gamma$ in the range 0.01 to 0.1.
This is repeated for each sample, using a new set of noises $\eta_{\bo{x}}(t)$.
The stationarity of the ensemble is checked by tracking ensemble averages of various observables, and this allows us to determine appropriate $t_{*}$. These times were $t_{*}=\mc{O}(10/\gamma)$, with some variation depending on parameters and $\sigma$.

When changing $\sigma$ to move between the CE and GCE, we keep the chemical potential $\mu$ constant for each temperature and interaction strength.
This assumption aids in obtaining a sequence of physically related intermediate ensembles. 
The value of $\mu$ is chosen so that the mean number of atoms $\langle N \rangle$ in the GCE  matches the CE value of $\wb{N}$. 
This  helps to avoid the possible competition between $\wb{N}$ and $\mu$ that was discussed in Sec.~\ref{TRANS}.

\subsection{Low temperature case}
\label{LOWT}

\begin{figure}
\begin{center}
\includegraphics[width=\columnwidth]{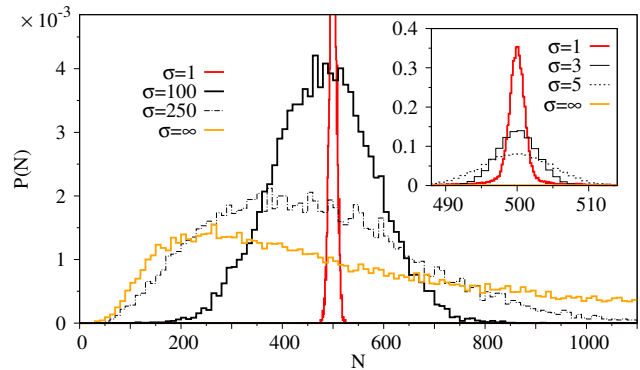}
\end{center}
\caption{
Probability distribution of the total number of atoms as the CE$\,\to\,$GCE parameter $\sigma$ is varied.
The inset shows the range of $\sigma$ values for which the properties of the ensemble are very close to an ideal canonical Bose gas.
Low temperature case, $T=0.341T_*$, $\overline{N}=500$.
}
\label{fig:lowT-hist-PN}
\end{figure}

Let us consider first a low temperature case in which the majority of atoms are in the condensate. This is the regime in which $P(N_0)$ or $P(N_{\rm ex})$ distributions have most commonly been described. For example \cite{Grossmann96,Weiss97,Wilkens97,Sinatra02,Svidzinsky06,Witkowska09,Witkowska10,Bienias11a,Heller13,Tarasov15} in the ideal gas, and \cite{Sinatra02,Carusotto03c,Bezett09b,Idziaszek09,Bisset09b,Witkowska10,Cockburn11a,Bienias11a,vanderWurff14} in the interacting.
One reason for its popularity is that it is accessible by the Bogoliubov approximation.

Fig.~\ref{fig:lowT-hist-PNex} shows the distribution of the number of excited atoms $N_{\rm ex}$ at $T=0.341\, T_*$ in the CE and GCE. The CE has  $\wb{N}=500$ and the ensemble is obtained using \eqn{SPGPEs} with $\sigma=1$. 
In the c-field description, $N_{\rm ex}=\sum_{k\neq0}|\wt{\psi}_{k}|^2 \Delta k$, and $\wt{\psi}$ is the Fourier transformed field
\eq{FT}{
\wt{\psi}_{k} = \frac{\sqrt{2\pi}}{L}\sum_{x} e^{-ikx} \psi_x.
}
The GCE is obtained using a simple SPGPE \eqn{SPGPE}, in the limit $\sigma\to\infty$.
As explained in Sec.~\ref{PROC}, the chemical potential is chosen so that the mean number of atoms 
$\langle N \rangle$ in the GCE matches the CE value of $\wb{N}=500$. This is $\mu=-2.135/L^2$ here.
The numerical histograms are compared to exact results which are obtained in Appendix~\ref{APP:EXACT}. We see that despite the not so large  shift from CE to GCE in this regime, the distribution tracks it in detail. The histogram is from $\mc{S}=2.5\times 10^4$ samples of the ensemble.		

Fig.~\ref{fig:lowT-hist-PN} shows the behavior of the probability distribution of the total atom number, $P(N)$, as $\sigma$ is varied through the transition ensembles.  This is for the same low temperature case. As expected, we move from an extremely broad distribution in the GCE, through a Gaussian (initially broad, later narrow) which converges to an extremely narrow distribution around $N=\wb{N}$.

\subsection{High temperature case}
\label{HIGHT}

\begin{figure}
\begin{center}
\includegraphics[width=\columnwidth]{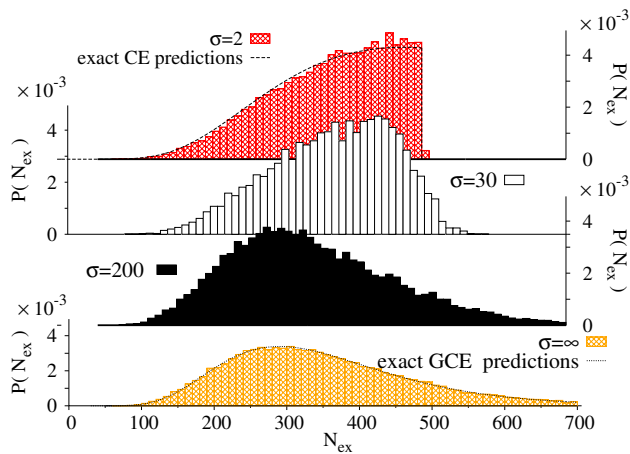}
\end{center}
\caption{
The progression of 
probability distributions of the number of excited atoms $N_{\rm ex}$ in the high temperature case of $T=1.365\, T_*$, $\overline{N}=500$.
From bottom to top, we have the  GCE (yellow), going through intermediate ensembles to the CE (upper, red).
Dotted and dashed lines correspond to the exact classical fields results \eqn{PNex-GCE} and \eqn{PNex-CE}, respectively. 
}
\label{fig:highT-hist-PNex}
\end{figure}

Distributions for the high temperature case have been reported for the ideal gas \cite{Grossmann96,Wilkens97,Weiss97,Svidzinsky06,Witkowska09,Witkowska10,Tarasov15} and interacting gas \cite{Carusotto03c,Bezett09b,Bisset09b,Cockburn11a,Bienias11a} in this regime. They behave very differently, though this has not been analyzed as much in the literature.

 Fig.~\ref{fig:highT-hist-PNex} shows the CE, GCE, and two intermediate ensembles for $T=1.365\, T_*$. 
 We set $\wb{N}=500$ and use $\mu=-32.789/L^2$ to have matching $\langle N\rangle=500$ in the GCE.
The match to exact CE and GCE results is ideal. Particularly notable is the reconstruction of the CE ``cliff'' in $P(N_{\rm ex})$ despite the total atom number not being hardwired into the simulation, and all values being at least in principle allowed. 

This is an unusual regime in a number of aspects. Apart from the presence of the sharp cliff, another interesting feature appears. Namely, 
the most commonly occurring values of the number of excited atoms are larger in the CE (and the $\sigma=30$ case) than in the GCE.
They are around 400 versus 300, respectively. This is rather counterintuitive compared to the usual impression that the GCE in the ideal gas allows much larger numbers of excited particles. What we observe here is a consequence of the strong restriction on allowable states that the CE (or low $\sigma$) condition imposes.

Further ideal gas results will appear as limiting cases in the later discussion and Figs.~\ref{fig:lowT-dN0/N0} to~\ref{fig:equiv-highT}, such as $P(N_0)$, mean values of $\langle N\rangle$, and fluctuations of $N$ and $N_0$.

\section{Interacting gas and transitional ensembles}
\label{INT}

\begin{figure}
\begin{center}
\includegraphics[width=\columnwidth]{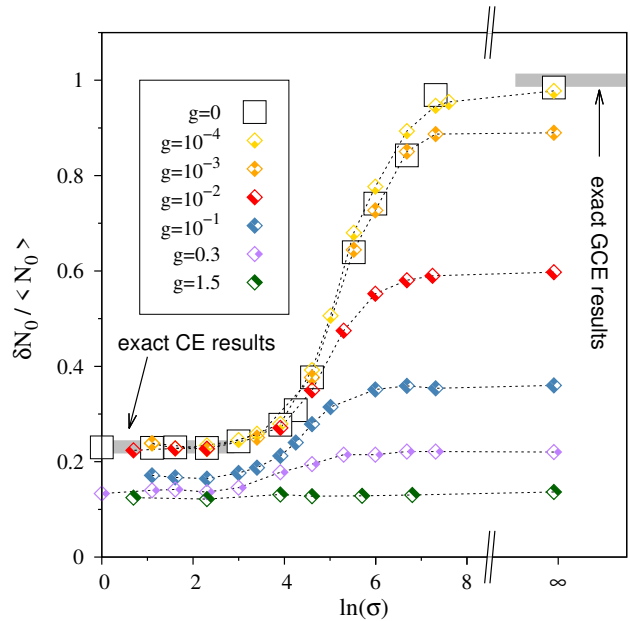}
\end{center}
\caption{
Relative variance of condensate atom number $N_0$ in the low temperature ($T=0.341\, T_*, \wb{N}=500$) regime as a function of the $\sigma$ parameter
for a wide spectrum of interaction values $g$ (in units of $1/L$). Gray lines are exact asymptotic results for the ideal gas system ($g=0$) in the CE and GCE. 
}
\label{fig:lowT-dN0/N0}
\end{figure}

Having verified that the method reproduces the expected ideal gas distributions exactly, we now turn to a more detailed analysis of  
the effect of nonzero interactions on the transitional ensembles.

Fig.~\ref{fig:lowT-dN0/N0} shows how the relative fluctuation of the number of condensate atoms
\eq{delN0}{
\frac{\delta N_0}{\langle N_0\rangle} = \frac{\sqrt{\langle N_0^2 \rangle - \langle N_0 \rangle^2}}{\langle N_0 \rangle}
}
changes with $\sigma$ and $g$ in the low temperature case $T=0.341\, T_*$. 
The behavior of this quantity when temperature, $N$ or interaction are changed has been studied extensively in the standard ensembles (CE,GCE,MCE) \cite{Grossmann96,Weiss97,Navez97,Grossmann97,Wilkens97,Sinatra02,Svidzinsky06,Idziaszek09,Witkowska10,Cockburn11a,Bienias11a,Bienias11b,Heller13,Weiss16}, but not the transition between ensembles or experimentally relevant intermediate cases.

\begin{figure}
\begin{center}
\includegraphics[width=\columnwidth]{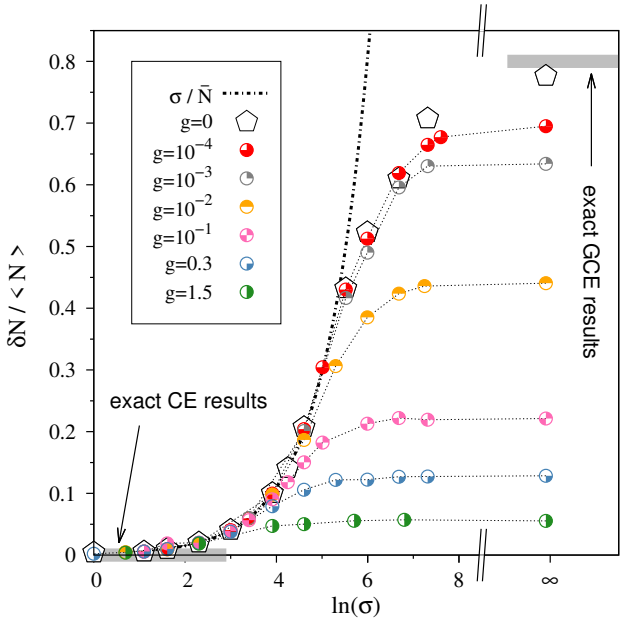}
\end{center}
\caption{  
Relative variance of total atom number $N$ in the low temperature ($T=0.341\, T_*, \wb{N}=500$) regime. All notation same as in Fig.~\ref{fig:lowT-dN0/N0}.
The dashed line $\delta N/\langle N\rangle = \sigma/\wb{N}$  is a naive estimation of the variance due to just the effect of the external parameter $\sigma$. 
}
\label{fig:lowT-dN/N}
\end{figure}

On the figure, the ideal gas case appears as hollow symbols, and unsurprisingly has the highest relative fluctuations. 
We see two plateau regions. The first, for  $\sigma\lesssim20$, in which there is no discernible difference from the CE. 
The size of this range is related to the width of typical features in the distribution of $P(N_0)$. 
When the allowable fluctuation in $N$ (which is $\sim\sigma$) becomes several times smaller, it will cease to visibly affect $P(N_0)$. For example, in Fig.~\ref{fig:lowT-hist-PNex} features in the distribution of $P(N_{\rm ex})$ have a width of $\mc{O}(50)$ atoms, and the same applies for $P(N_0)$ in the CE.

The second plateau area for $\sigma\gtrsim 1000$ displays the same magnitude of fluctuations of $N_0$ as in the GCE. 
Note that this is a point where $\sigma\sim\wb{N}$, and indeed we would not expect the Gaussian narrowing caused by $\sigma$ to affect much if it is significantly broader than the natural size of $N_0$ fluctuations in the GCE.

As interaction is raised, initially only the fluctuations in the GCE are affected because they are large. This starts for quite small interaction strengths.  As interaction grows, the GCE-like region expands somewhat to lower values of $\sigma$.
Eventually, though, for strong-enough interaction, the fluctuations of $N_0$ begin to reduce also in the CE, somewhat unexpectedly. 

Figure~\ref{fig:lowT-dN/N} shows the relative fluctuation of the \emph{total} number of atoms, $N$.
This has not been studied so much in the standard ensembles, primarily because not much happens in those cases (e.g. in the CE or MCE). 
For intermediate ensembles, we also see the two plateau regions in the limits of $\sigma$ and strong reductions in fluctuation with increasing $g$.

A comparison of $\delta N/\langle N\rangle$ with the naively expected effect of only the Gaussian narrowing (which is $\sigma/\wb{N}$) in Fig.~\ref{fig:lowT-dN/N} shows that the total particle fluctuations track this estimate faithfully from small $\sigma$ up to $\sigma\approx150$. This agrees with \eqn{deltaN}. Note that $\delta^{GCE}N \approx 400$, 
and the center of the transition between $\sigma$-limited behavior and natural GCE behavior in Fig.~\ref{fig:lowT-dN/N} is also around this value of $\sigma$. 

\begin{figure}
\begin{center}
\includegraphics[width=\columnwidth]{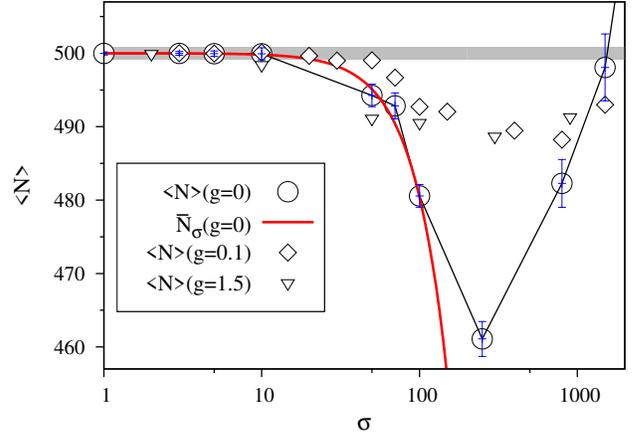}
\end{center}
\caption{
Mean number of atoms as the fluctuation control parameter $\sigma$ changes. Various interaction strengths, and the ideal gas case, are shown. Low temperature case, $T=0.341\, T_*$, $\wb{N}=500$. 
The red line shows the simple estimate \eqn{Nmu} for the ideal gas. 
}
\label{fig:N}
\end{figure}

To see in more detail what goes on in the transitional ensembles, Fig.~\ref{fig:N} shows the mean atom number as a function of $\sigma$. 
There is some (mostly minor) variation despite $\mu$ being chosen to match CE and GCE mean atom numbers in the two limiting cases.
At low values of $\sigma$, the ideal gas behaves as predicted by \eqn{Nmu} (red line).
Overall, there is a dip at intermediate $\sigma$.
This comes about because the GCE distribution of $N$ has a positive skewness (long tail at high $N$).
The tail is more strongly suppressed by the Gaussian multiplier in \eqn{PGCEs} than the low $N$ part of the distribution, because the latter is closer to the mean.

\begin{figure}
\begin{center}
\includegraphics[width=\columnwidth]{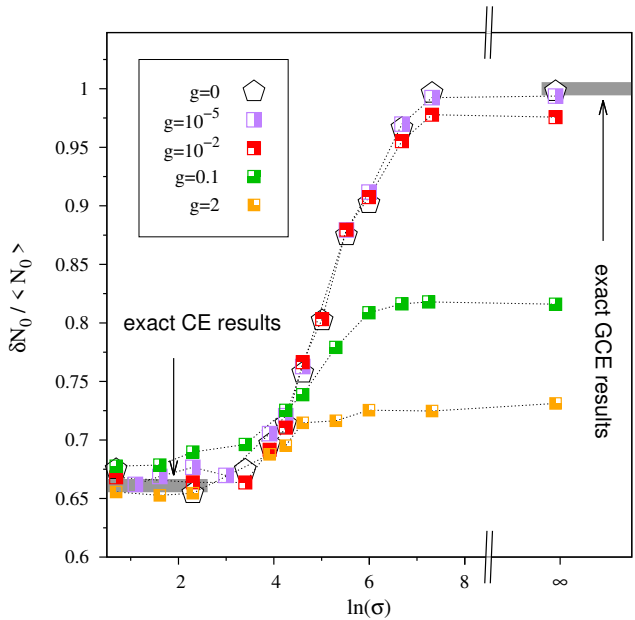}
\end{center}
\caption{
Relative variance of condensate atom number $N_0$ in the high temperature ($T=1.365\, T_*, \wb{N}=500$) regime. Notation is the same as in Fig.~\ref{fig:lowT-dN0/N0}.
}
\label{fig:highT-dN0/N0}
\end{figure}

Finally, Fig.~\ref{fig:highT-dN0/N0} shows the relative fluctuations of the ground state (``condensate'')  occupation for the high temperature case $T=1.365\, T_*$. 
The relative fluctuations are large even in the CE. 
Note that this occupation is still appreciable (in the range 0-300) despite $T>T_*$, since we are considering a mesoscopic system, not one in the limit of $N\to\infty$. 
The usual plateau behaviors seen before in Fig.~\ref{fig:lowT-dN0/N0} are also present. Differently from low $T$, it takes a rather strong interaction $g\ge0.01$  (units of $1/L$) to invoke a response in the relative fluctuations. Moreover it is difficult to bring the fluctuations down to the CE level by interactions alone, despite the CE value of $\delta N_0/N=0.66$ being very high. 
e.g. $g=2$ is still not fully sufficient.

\section{Equivalence of ensembles}
\label{EQUIV}

\begin{figure}
\begin{center}
\includegraphics[width=\columnwidth]{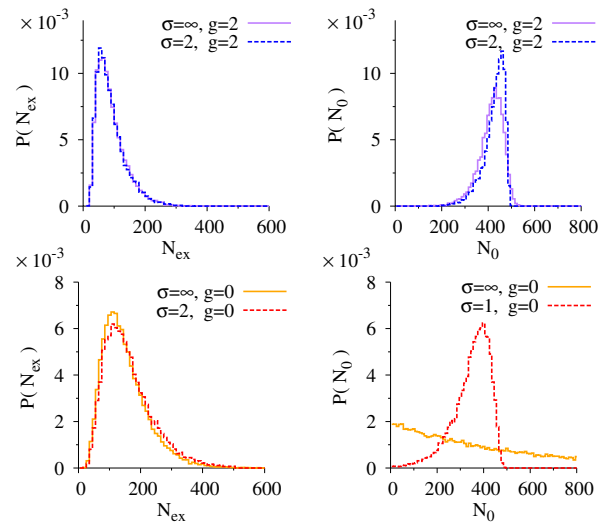}
\end{center}
\caption{
A demonstration of the equivalence of ensembles as interaction rises.
Low temperature ($T=0.341\, T_*, \wb{N}=500$) probability distributions of $N_{\rm ex}$ (left) and $N_0$ (right) in the case of large interaction, $g=2$ -- top panel, 
in comparison to the probability distributions in the non-interacting atom case -- lower panel.
}
\label{fig:equiv-lowT}
\end{figure}

Let us now look in some more detail at the matter of the equivalence of ensembles as interaction is increased. 
In most studies, the relative fluctuation of $N_0$ is the quantity that has been investigated in this context. For equivalence, we expect the CE and GCE values to be equal
(or ideally, to see a horizontal line across all $\sigma$ values). 
This is indeed what is seen in Fig.~\ref{fig:lowT-dN0/N0} for the highest values of $g$ (e.g. it is close for $g=0.3$ and truly equal at $g=1.5$). A very good match for our parameters occurs only once the CE value (and the rest) have fallen below the ideal gas value due to interactions. At this stage we have insufficient information to state whether this is a general feature.

The detailed behavior of the distributions is shown for this low $T$ case in Fig.~\ref{fig:equiv-lowT}, in which each panel compares the CE and GCE distributions.
The lower panels show the ideal gas, and apart from the huge discrepancy in $P(N_0)$, we see that $P(N_{\rm ex})$ also differs. At the strong interaction of $g=2$, however, both distributions have become very close. A small difference in $P(N_0)$ remains, though it is of a size that would often be inconsequential operationally.

The above plots quantitatively validate many existing intuitions about ensemble equivalence.

\begin{figure}
\begin{center}
\includegraphics[width=\columnwidth]{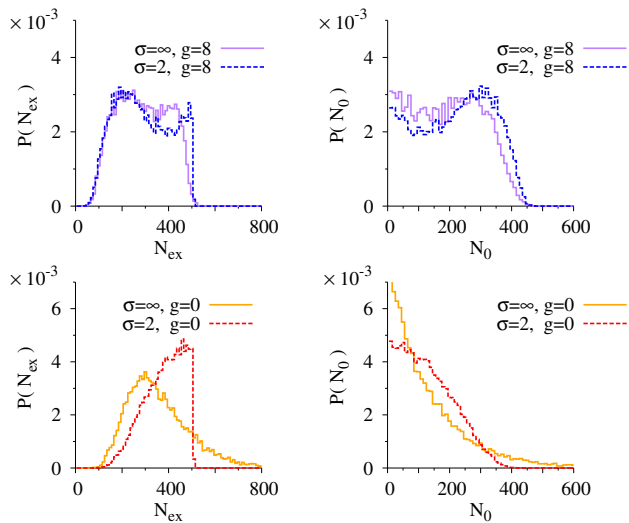}
\end{center}
\caption{
Approach to equivalence of ensembles in the high  temperature ($T=1.365\, T_*, \wb{N}=500$) case. Notation as in Fig.~\ref{fig:equiv-lowT}.
Probability distributions of $N_{\rm ex}$ (left) and $N_0$ (right) in the case of large interaction, $g=8$ -- top panel, 
in comparison to the probability distributions in the non-interacting atom case -- lower panel.
Note the robust reproduction of the cliff in $P(N_{\rm ex})$ near $N_{\rm ex}=\wb{N}$.
}
\label{fig:equiv-highT}
\end{figure}

The system is somewhat more resistant to ensemble equivalence at the higher temperatures above $T_*$ in Fig.~\ref{fig:highT-dN0/N0}. 
The behavior of the distributions (in which we expect the appearance of the cliff) is shown for this case in Fig.~\ref{fig:equiv-highT}.
In the ideal gas  the CE and GCE distributions are dissimilar for both $N_{\rm ex}$ and $N_0$, though the difference for $N_0$ is nowhere near as great as at lower temperatures. 
In the end though, at the high interaction value of $g=8$  shown (larger than in Fig.~\ref{fig:highT-dN0/N0}), the shape of the CE and GCE distributions in the upper panels have become very close. The conclusion is that ensemble equivalence has been restored by interactions also here. 
This is despite the complicated form of the CE/GCE distribution itself.

Of particular note is the fact that the ``cliff'' near $N_{\rm ex}=500$ is also present in the GCE!
In the ideal gas, this feature was due exclusively to the property $P(N_0)=P(\wb{N}-N_{\rm, ex})$ caused by the hard-wired external constraint $\wb{N}=500$ in the CE, and was rather fragile with respect to a change of the ensemble. 
For example at $\sigma=30$, the cliff has already practically disappeared in Fig.~\ref{fig:lowT-hist-PNex}, while other quantities such as $\delta N_0/\langle N_0\rangle$ are more robust, and have hardly budged from their CE value.
However, in Fig.~\ref{fig:equiv-highT} interactions impose the cliff again in the GCE where there is absolutely no explicit constraint on particle number.
This quantitatively validates ensemble equivalence at high temperatures, including the preservation of features that appear fragile in the ideal gas.

\section{Conclusions}
\label{CONCLUSIONS}

We have derived extended SPGPE-like equations that generate a canonical ensemble (rather than grand canonical) as the stationary state. This occurs in the limit of small $\sigma$, which is a control parameter for the allowed fluctuations of the total atom number 
\eq{dn=s}{
\delta N\approx\sigma.
}
The equations are easily adaptable to arbitrary external potentials and nonlocal interactions.
 A major added benefit is the possibility to readily generate a whole range of intermediate ensembles between canonical and grand canonical. The relationship \eqn{dn=s} makes it quite simple to use it to match the true experimental variability of atom number in a run with many experimental realizations.

We have tested the ensembles produced (Sec.~\ref{IDGAS}), and also shown their utility for studying the transition to the canonical ensemble and the onset of ensemble equivalence as interaction grows. We have also drawn attention to the unusual behavior of canonical ensembles with low numbers of atoms: the appearance of a cliff in the distribution of atoms in excited modes, and its retention also in the grand canonical ensemble when interactions are sufficiently high. 

It is hoped that the method will be useful for the study of canonical ensembles, other experimentally obtained ensembles that do not fit neatly into the CE/GCE categorization, as well as for the study of ensemble equivalence and other related phenomena. 
 Here, we have quantitatively validated the ensemble equivalence scenario and shown the details of how it gradually appears with growing $g$ (Figs.~\ref{fig:lowT-dN0/N0},\ref{fig:lowT-dN/N},\ref{fig:highT-dN0/N0}-\ref{fig:equiv-highT}). Interestingly and importantly, we find that even canonical ensemble features such as the ``cliff'' that are fragile to a weakening of the canonical ensemble constraint in the ideal gas can nevertheless be robustly reproduced when interactions become strong. 

Importantly, the equations  should be readily applicable to multiple components. One $-\gamma(N_j-\wb{N}_j)\psi_j(\bo{x})/\sigma_{j}^2$ term
 can be added to the evolution equation $d\psi_j(\bo{x})/dt$ for each component $j$.
 This is much preferable to trying to set the magnetization or relative populations of different components using Lagrange multipliers $\mu_j$ because the latter only set the \emph{mean} particle number and may allow very large relative fluctuations from shot to shot. This is a work in progress.

Looking further ahead, the equations presented here should be capable of producing ensembles of attractive gases with $g<0$, something that absolutely cannot be stably treated using the standard SGPE. Moreover, they should also be easily extensible to the case of long range interactions,  another situation when ensemble equivalence is known to be broken \cite{Touchette15}.

Recent experiments have demonstrated atom number measurements well below atomic shot noise using dispersive imaging \cite{Gajdacz16,Kristensen16}, 
which suggests that high precision studies of ensemble equivalence could be carried out with present technology.

\acknowledgments
 This work was supported by the National Science Centre grant No. 2012/07/E/ST2/01389. 

\bibliography{cfields}

\appendix


\section{The case of nonuniform $\gamma(\bo{x})$}
\label{APP:GAMMA}

If we do not assume \eqn{gamconst}, then instead of \eqn{FPE-1} one obtains the following even more cumbersome form of the FPE:
\begin{widetext}
\eqa{FPE-A1}{
\frac{\partial P_{\sigma}}{\partial t} &=&
2P_{\sigma}\sum_{\bo{x}}\mc{P}_{\bo{x}\bo{x}}\Big[\left(2g|\psi_{\bo{x}}|^2-\mu\right)\gamma_{\bo{x}}-(N-\wb{N}){\rm Re}[K_{\bo{x}}]\Big]
-2P_{\sigma}\sum_{\bo{x}\bo{z}}\mc{P}_{\bo{x}\bo{z}}\gamma_{\bo{z}}\mc{P}_{\bo{z}\bo{x}}\left[2g|\psi_{\bo{x}}|^2-\mu+\frac{T(N-\wb{N})}{\sigma^2}\right]
\nonu\\
&&+P_{\sigma}\frac{N-\wb{N}}{T}\left[\frac{T(N-\wb{N})}{\sigma^2}-\mu\right]\sum_{\bo{x}\bo{y}}dV\psi_{\bo{y}}\psi_{\bo{x}}^*\left\{\frac{2T}{\sigma^2}\sum_{\bo{z}}\mc{P}_{\bo{x}\bo{z}}\gamma_{\bo{z}}\mc{P}_{\bo{z}\bo{y}}+\mc{P}_{\bo{x}\bo{y}}K_{\bo{y}} + \mc{P}_{\bo{x}\bo{y}}K_{\bo{x}}^*\right\}
\nonu\\
&&-2P_{\sigma}\sum_{\bo{x}\bo{y}}dV\psi_{\bo{x}}\psi_{\bo{y}}^*\left\{\mc{P}_{\bo{x}\bo{y}}{\rm Re}[K_{\bo{y}}]+\frac{T}{\sigma^2}\sum_{\bo{z}}\mc{P}_{\bo{x}\bo{z}}\gamma_{\bo{z}}\mc{P}_{\bo{z}\bo{y}}\right\}
\\
&&+P_{\sigma}\frac{N-\wb{N}}{T}\sum_{\bo{x}\bo{y}}dV\psi_{\bo{x}}\psi_{\bo{y}}^*\left[g\psi_{\bo{y}}|\psi_{\bo{y}}|^2+\sum_{\bo{z}'}\psi_{\bo{z}'}H_{\bo{y}\bo{z}'}^{\rm sp}\right]\left\{\mc{P}_{\bo{x}\bo{y}}\left(K_{\bo{x}}^*-\frac{\gamma_{\bo{y}}+i}{\sigma^2}\right)+ \frac{2T}{\sigma^2}\sum_{\bo{z}} \mc{P}_{\bo{x}\bo{z}}\gamma_{\bo{z}}\mc{P}_{\bo{z}\bo{y}}\right\} + {\rm c.c.}
\nonu\\
&&+P_{\sigma}\frac{dV}{T}\sum_{\bo{x}\bo{y}}\left[2\sum_{\bo{z}}\mc{P}_{\bo{x}\bo{z}}\gamma_{\bo{z}}\mc{P}_{\bo{z}\bo{y}}-\mc{P}_{\bo{x}\bo{y}}\left(\gamma_{\bo{x}}+\gamma_{\bo{y}}\right)\right]
\Bigg\{\psi_{\bo{y}}\psi_{\bo{x}}^*\left[(g|\psi_{\bo{y}}|^2-\mu)(g|\psi_{\bo{x}}|^2-\mu)-\frac{T\mu(N-\wb{N})}{\sigma^2}\right] \nonu\\
&&\hspace*{5em}
+ \sum_{\bo{z}'\bo{z}''}H_{\bo{y}\bo{z}'}^{\rm sp}H_{\bo{z}''\bo{x}}^{\rm sp} \psi_{\bo{z}'}\psi_{\bo{z}''}^* 
+\psi_{\bo{x}}^*\sum_{\bo{z}'}H_{\bo{y}\bo{z}'}^{\rm sp}\psi_{\bo{z}'}(g|\psi_{\bo{y}}|^2-\mu)
+\psi_{\bo{y}}\sum_{\bo{z}'}H_{\bo{z}'\bo{x}}^{\rm sp}\psi_{\bo{z}'}^*(g|\psi_{\bo{x}}|^2-\mu)
\Bigg\}.
\nonu
}
\end{widetext}
We omitted the $\psi$ dependence of $N(\psi)$ for a minor improvement in brevity. 
Due to the fact that in many terms there is no summation index that involves only $\mc{P}$ and $\psi$, one cannot apply \eqn{Pxypsix} in all necessary cases as was done for a constant $\gamma$.

A special but very common case is when the projection is made implicitly by the numerical lattice as done in the plain SGPE approach (rather than the SPGPE). Then, $\mc{P}_{\bo{x}\bo{y}}=\delta_{\bo{x}\bo{y}}$ and all the inconvenient features of \eqn{FPE-A1} abate. One obtains
\eqa{FPE-A2latt}{
\frac{\partial P_{\sigma}}{\partial t} &=& 
2P_{\sigma}\sum_{\bo{x}}\left[{\rm Re}[K_{\bo{x}}] + \frac{\gamma_{\bo{x}} T}{\sigma^2}\right]
\Bigg\{
-(N-\wb{N}+|\psi_{\bo{x}}|^2dV)\nonu\\
&&\hspace*{-2em}
+\frac{N-\wb{N}}{T}\left[g|\psi_{\bo{x}}|^2-\mu+\frac{T(N-\wb{N})}{\sigma^2}\right] |\psi_{\bo{x}}|^2 dV
\Bigg\}\\
&&\hspace*{-2em}+2P_{\sigma}\frac{N-\wb{N}}{T}\sum_{\bo{x}}dV{\rm Re}\left[\left(K_{\bo{x}}^*+\frac{T\gamma_{\bo{x}}}{\sigma^2}\right)\psi_{\bo{x}}^*\sum_{\bo{z}}H_{\bo{x}\bo{z}}^{\rm sp}\psi_{\bo{z}}\right],
\nonu
}
which is similar in complexity to \eqn{FPE-2}.
This still complicated expression can be made zero with the simple choice
\eq{K=latt}{
K_{\bo{x}} = -\frac{\gamma_{\bo{x}} T}{\sigma^2}.
}
in full space-dependent analogy to \eqn{K=}.

Now, if we return to the general projected case,  the typical situation is that $\gamma(\bo{x})$ is slowly varying in space compared to the highest energy modes in $\mc{C}$.
These produce features of length $\lambda_E$.  On the other hand, $\mc{P}_{\bo{x}\bo{y}}$ is typically close to diagonal with a width given by the length scale of the highest-energy components.
This means that it decays to zero on length scales of the order of $\lambda_E$. 
Thus, as long as $\gamma(\bo{x})$ varies slowly in the region of space around $\bo{x}$ compared to $\mc{P}_{\bo{x}\bo{y}}$, one will have
\eq{approx}{
\mc{P}_{\bo{x}\bo{y}}\gamma_{\bo{x}} \approx \mc{P}_{\bo{x}\bo{y}}\gamma_{\bo{y}}.
} 
This condition allows us to put the troublesome terms in \eqn{FPE-A1} involving $\gamma$ into a form in which the projection property of the field, \eqn{Pxypsix}, can be applied. For example, one has
\eqs{help}{
\sum_{\bo{z}}\mc{P}_{\bo{x}\bo{z}}\gamma_{\bo{z}}\mc{P}_{\bo{z}\bo{y}} \to \mc{P}_{\bo{x}\bo{y}}\gamma_{\bo{x}}\\
\left[2\sum_{\bo{z}}\mc{P}_{\bo{x}\bo{z}}\gamma_{\bo{z}}\mc{P}_{\bo{z}\bo{y}}-\mc{P}_{\bo{x}\bo{y}}\left(\gamma_{\bo{x}}+\gamma_{\bo{y}}\right)\right] \to 0.
}
However, there are some remaining (also troublesome) terms in \eqn{FPE-A1} which involve $K_{\bo{x}}$ not $\gamma_{\bo{x}}$. Judging by the earlier result \eqn{K=latt} for a special case, the rate at which $K$ will vary spatially is 
similar to that of $\gamma$. So, let us also provisionally assume the same slowly varying property for $K$, and check its consistency later. 
This assumption lets us apply
\eq{approxK}{
\mc{P}_{\bo{x}\bo{y}}K_{\bo{x}} \approx \mc{P}_{\bo{x}\bo{y}}K_{\bo{y}}.
} 
Conditions \eqn{approx}-\eqn{approxK} lead to much simplification in \eqn{FPE-A1}:
\eqa{FPE-A2gen}{
\frac{\partial P_{\sigma}}{\partial t} &=& 
2P_{\sigma}\sum_{\bo{x}}\left[{\rm Re}[K_{\bo{x}}] + \frac{\gamma_{\bo{x}} T}{\sigma^2}\right]
\Bigg\{
(\wb{N}-N)\mc{P}_{\bo{x}\bo{x}}-|\psi_{\bo{x}}|^2dV\nonu\\
&&\hspace*{-2em}
+\frac{N-\wb{N}}{T}\left[g|\psi_{\bo{x}}|^2-\mu+\frac{T(N-\wb{N})}{\sigma^2}\right] |\psi_{\bo{x}}|^2 dV
\Bigg\}\\
&&\hspace*{-2em}+2P_{\sigma}\frac{N-\wb{N}}{T}\sum_{\bo{x}}dV{\rm Re}\left[\left(K_{\bo{x}}^*+\frac{T\gamma_{\bo{x}}}{\sigma^2}\right)\psi_{\bo{x}}^*\sum_{\bo{z}}H_{\bo{x}\bo{z}}^{\rm sp}\psi_{\bo{z}}\right.
\nonu
}
The above equation becomes stationary using the same simple expression \eqn{K=latt} as for the unprojected case. 
This confirms the  validity of the condition \eqn{approxK} once \eqn{approx} is assumed.

Substituting \eqn{K=latt} into the general postulated stochastic equation \eqn{SPGPEmod} gives us the most general transition SPGPE  \eqn{SPGPEs-gen}.

\section{Some exact results for the ideal gas}
\label{APP:EXACT}

We follow the same approach as \cite{Witkowska09} used for the 1d trapped gas, but adapt the procedure 
for the doubly degenerate levels that occur in the uniform gas. 

We have plane wave modes $k_j$ with energies 
\eq{vej}{
\ve_j = \frac{\hbar^2k_j^2}{2m},
}
occupied by $n_j$ bosons, and temperature set by $\beta=1/k_BT$. The total energy of a state is $E=\sum_j\ve_jn_j$, and the number of atoms is $N=\sum_jn_j$. 

\subsection{Canonical ensemble }

The fuller version of  (1) from \cite{Witkowska09} that also includes the canonical ensemble constraint $N=\wb{N}$ is
\eq{full}{
P(N_{\rm ex}) = \left[ \prod_j \sum_{n_j=0}^{\infty} e^{-\beta\ve_jn_j}\right] \Mbig{\delta}_{N_{\rm ex},\,\sum_{j\neq0} n_j} \ \Mbig{\delta}_{\wb{N},\,\sum_j n_j}.
}
$j$ enumerates modes over all integers, including the ground state $j=0$ which is the only nondegenerate mode and sets the energy zero: $\ve_0=0$. 
Normalization of $P$ is ignored. 
Combining the two deltas immediately implies the obvious 
$n_0 = \wb{N}-N_{\rm ex}$,
and the necessity of the ``cliff'', i.e. $P(N_{\rm ex}>\wb{N})=0$. 

Substituting 
\eq{delta}{
\delta_{a,b} = \frac{1}{2\pi}\int_0^{2\pi} e^{ix(a-b)} dx
}
into \eqn{full} we get
\eq{eq1}{
P(N_{\rm ex}) = \frac{1}{2\pi} \int_0^{2\pi}\!dx\ e^{ixN_{\rm ex}} \prod_{j\neq0} \left[
\sum_{n_j=0}^{\infty} e^{-n_j(\beta\ve_j+ix)}
\right]
}
The sum has the form $\sum_{n=0}^{\infty} f^n = \frac{1}{1-f}$ where $f=e^{-\beta\ve_j-ix}$. So
\eq{eq2}{
P(N_{\rm ex}) = \frac{1}{2\pi} \int_0^{2\pi}\!dx\ e^{ixN_{\rm ex}} \prod_{j\neq0} 
\frac{1}{1-e^{-\beta\ve_j-ix}}
}
We now have a right-hand contour on  the unit circle in the variable $z=e^{ix}$. In particular, $dz=izdx$, so
\eq{eq3}{
P(N_{\rm ex}) = \frac{-i}{2\pi} \oint\!dz\ z^{N_{\rm ex}-1} \prod_{j\neq0} 
\frac{z}{z-e^{-\beta\ve_j}}.
}
The poles (doubly degenerate) are  at locations
\eq{ak}{
a_m = \exp[-\beta\ve_m], \qquad \forall\ve_m>0
}
an all within the contour because $\beta\ve_k>0$. Note that  $m\ge0$ now counts only energy levels, not modes, and we denote $m=0$ to be the ground state.
With the help of the Cauchy residue theorem, the result is
\eqs{eq4}{
P(N_{\rm ex}) &=& \sum_{m>0} {\rm Res}\left[z^{N_{\rm ex}-1} \prod_{j\neq0} \frac{z}{z-e^{-\beta\ve_j}}\,,\ a_m\right] \\
&=&a_m^{N_{\rm ex}} \left[1+N_{\rm ex} + 2\sum_{m''\neq m,0}\frac{1}{1-a_m/a_{m''}}\right]\nonu\\
&&\qquad\times\prod_{m'\neq m,0} \left(\frac{1}{1-a_{m'}/a_m}\right)^2.
}
Upon substitution, the sum is now over excited energy levels, and the end result looks like this:
\begin{widetext}
\eq{PNex-CE-Q}{
P(N_{\rm ex}) = \ifsplit{
\sum_{\ve_k>0} e^{-\beta\ve_k N_{\rm ex}} \left[1+N_{\rm ex}+2\sum_{\ve_{k'}\neq 0,\ve_k}\frac{1}{1-e^{-\beta(\ve_k-\ve_{k'})}}\right]\prod_{\ve_l\neq 0,\ve_k}\left(\frac{1}{1-e^{-\beta(\ve_{l}-\ve_k)}}\right)^2.
& \text{if $0\le N_{\rm ex}\le \wb{N}$}\\
0 & \text{if $N_{\rm ex}> \wb{N}$}
}
}
\end{widetext}

\subsection{Classical field expression}
In the c-field approximation, again analogously to \cite{Witkowska09}, the expression corresponding to \eqn{full} is
\eqa{Cfull}{
P_c(N_{\rm ex}) &=& \left[ \prod_{j\in\mc{C}} \int \frac{d^2\alpha_j}{\pi}  e^{-\beta\ve_j|\alpha_j|^2}\right] \\
&&\times\delta\left( N_{\rm ex}-\sum_{j\in\mc{C},\neq0} |\alpha_j|^2\right)\delta\left( \wb{N}-\sum_j |\alpha_j|^2\right)\nonu
}
with Dirac deltas and mode amplitudes $\alpha_j$.
Here also the two deltas give a deterministic condition 
$\delta\left(|\alpha_0|^2 - (\wb{N}-N_{\rm ex})\,\right)$
on the ground state amplitude, and the ``cliff'' is present as well. 
Moving on to the evaluation of this expression for degenerate states, we use \eqn{delta} again and find
\eq{Ceq1}{
P_c(N_{\rm ex}) = 
\frac{1}{2\pi} \int_0^{2\pi}\!dx\ e^{ixN_{\rm ex}} \prod_{j\in\mc{C},\neq0} \left[
\int \frac{d^2\alpha_j}{\pi}  e^{-|\alpha_j|^2(\beta\ve_j+ix)}
\right].
}
The integrals are easily done, giving
\eq{Ceq2}{
P_c(N_{\rm ex}) = 
\frac{1}{2\pi} \int_0^{2\pi}\!dx\ e^{ixN_{\rm ex}} \prod_{j\in\mc{C},\neq0} \frac{1}{\beta\ve_j+ix}
}
Changing to contour variable $z$, 
\eqa{Ceq3}{
P_c(N_{\rm ex}) &=& \frac{-i}{2\pi} \oint\!dz\ z^{N_{\rm ex}-1} \prod_{j\in\mc{C},\neq0} 
\frac{1}{\beta\ve_j+\log z}\\
&=& \sum_{m>0} {\rm Res}\left[z^{N_{\rm ex}-1} \prod_{j\in\mc{C},\neq0} 
\frac{1}{\beta\ve_j+\log z}, a_m\right].\nonu
}
The poles are at the same locations \eqn{ak} as before in the quantum case, with the same degeneracy, so that  
evaluation of the residues leads to
\begin{widetext}
\eq{PNex-CE}{
P_c(N_{\rm ex}) = \ifsplit{
\sum_{\ve_k>0,\in\mc{C}} e^{-\beta\ve_k N_{\rm ex}} \left[N_{\rm ex}+2\sum_{\ve_{k'}\neq 0,\ve_k}\frac{1}{\beta(\ve_k-\ve_{k'})}\right]\prod_{\ve_l\in\mc{C},\neq 0,\neq\ve_k}\left(\frac{1}{\beta(\ve_{l}-\ve_k)}\right)^2.
& \text{if $0\le N_{\rm ex}\le \wb{N}$}\\
0 & \text{if $N_{\rm ex}> \wb{N}$}
}
}
\end{widetext}
This sum is overall similar to the quantum one \eqn{PNex-CE-Q}, replacing the Bose-Einstein occupations with Rayleigh-Jeans, working with only the states in the subspace $\mc{C}$, and without the extra $+1$ term in the prefactor.

\subsection{Grand canonical ensemble}
Consider now the GCE in the quantum case. Here, weights for states are
\eq{weights}{
e^{-\beta(\ve_j-\mu)n_j},
}
with $\mu<0$.
We have, rather similarly to \eqn{full}, that 
\eq{fullgceNex-a}{
P(N_{\rm ex}) = \prod_j \left[ \sum_{n_j=0}^{\infty} e^{-\beta(\ve_j-\mu)n_j}\right] \Mbig{\delta}_{N_{\rm ex},\,\sum_{j\neq0} n_j}.
}
with no constraint on $N$.
In principle the $j=0$ state enters above, but only as a prefactor $\sum_{n_0=0}^{\infty} e^{\beta\mu n_0} = 1/(1-e^{\beta\mu})$ that 
can be incorporated into the normalization. Hence $\prod_j \to \prod_{j\neq0}$ and the expression for $P(N_{\rm ex})$ differs from 
the expression \eqn{full} for the canonical ensemble only by the replacement $\ve_j\to(\ve_j-\mu)$. Proceeding as before, one obtains
\begin{widetext}
\eq{PNex-GCE-Q}{
P(N_{\rm ex}) = 
\sum_{\ve_k>0} e^{-\beta(\ve_k-\mu) N_{\rm ex}} \left[1+N_{\rm ex}+2\sum_{\ve_{k'}\neq 0,\ve_k}\frac{1}{1-e^{-\beta(\ve_k-\ve_{k'})}}\right]\prod_{\ve_l\neq 0,\ve_k}\left(\frac{1}{1-e^{-\beta(\ve_{l}-\ve_k)}}\right)^2,
}
and for the c-field case:
\eq{PNex-GCE}{
P_c(N_{\rm ex}) = \sum_{\ve_k>0,\in{\mc{C}}} e^{-\beta(\ve_k-\mu) N_{\rm ex}} \left[N_{\rm ex}+2\sum_{\ve_{k'}\neq 0,\ve_k}\frac{1}{\beta(\ve_k-\ve_{k'})}\right]\prod_{\ve_l\in\mc{C}, \neq 0,\neq\ve_k}\left(\frac{1}{\beta(\ve_{l}-\ve_k)}\right)^2.
}
\end{widetext}

\end{document}